\documentclass[twoside]{article}

\usepackage{PRIMEarxiv}

\usepackage[utf8]{inputenc} 
\usepackage[T1]{fontenc}    
\usepackage{hyperref}       
\usepackage{url}            
\usepackage{booktabs}       
\usepackage{amsfonts}       
\usepackage{nicefrac}       
\usepackage{microtype}      
\usepackage{lipsum}
\usepackage{fancyhdr}       
\usepackage{graphicx}       
\usepackage{soul}
\usepackage{times}

\usepackage[utf8]{inputenc}
\usepackage[english]{babel}
\usepackage{algorithmic}
\usepackage{amsthm}
\usepackage{makecell}
\usepackage[numbers]{natbib}
\usepackage{multicol}
\usepackage{array}
\usepackage{multirow}
\usepackage{graphicx}
\usepackage{amsmath}
\usepackage{amssymb}
\usepackage[T1]{fontenc}
\usepackage[linesnumbered,lined,ruled,commentsnumbered]{algorithm2e}
\usepackage{comment}
\newcolumntype{P}[1]{>{\centering\arraybackslash}p{#1}}
\usepackage[small]{caption}
\usepackage{graphicx}
\usepackage{tabularx}
\usepackage{subcaption}
\usepackage{mathtools}
\usepackage{blkarray}
\usepackage[dvipsnames]{xcolor}
\usepackage[utf8]{inputenc}
\usepackage[english]{babel}

\DeclareMathOperator*{\argmax}{arg\,max}

\graphicspath{{media/}}     

\title{Heterogeneous Graph Attention Networks for Learning Diverse Communication
}

\author{
  Esmaeil Seraj$^{1,*}$, Zheyuan Wang$^{1,*}$, Rohan Paleja$^{1,*}$, Matthew Sklar$^{1}$, Anirudh Patel$^2$, Matthew Gombolay$^1$  \\
  $^1$Institute for Robotics \& Intelligent Machines, Georgia Institute of Technology, Atlanta, GA, USA, 30332--0250 \\ %
  $^2$Sandia National Laboratory, Albuquerque, NM, USA, 87185 \\
  $^*$ These authors contributed equally to this work. \\
  \texttt{\{eseraj3, pjohnwang, rpaleja3, msklar3\}@gatech.edu}, \texttt{matthew.gombolay@cc.gatech.edu} \\
  \texttt{anipate@sandia.gov}
}

\begin{document}
\maketitle

\begin{abstract}
Multi-agent \textcolor{black}{teaming achieves better performance} when there is communication among participating agents allowing them to coordinate their actions for maximizing shared utility. However, \textcolor{black}{when collaborating a team of agents with different action and observation spaces,} information sharing \textcolor{black}{is not straightforward and requires customized communication protocols, depending on sender and receiver types. Without properly modeling such heterogeneity in agents,} communication becomes less helpful and could even deteriorate the multi-agent cooperation performance. We propose heterogeneous graph attention networks, called HetNet, to learn efficient and diverse communication models for \textcolor{black}{coordinating heterogeneous agents towards accomplishing tasks that are of collaborative nature. We propose a Multi-Agent Heterogeneous Actor-Critic (MAHAC) learning paradigm to obtain collaborative per-class policies and effective communication protocols for composite robot teams. Our proposed framework is evaluated} against multiple baselines in a complex environment in which agents of different types must communicate and cooperate to satisfy the objectives. \textcolor{black}{Experimental results show that HetNet outperforms the baselines in learning sophisticated multi-agent communication protocols by achieving $\sim$10\% improvements in performance metrics}.
\end{abstract}


\section{Introduction}
\label{sec:Introduction}
\noindent Information sharing is key in building team cognition, and enables agents to cooperate to successfully achieve shared goals~\cite{Salas1992TowardAU, seraj2021adaptive}. Not only is the communication protocol essential to developing shared mission goals, \textcolor{black}{communication between agents must be} targeted and efficient. High-performing human teams utilize communication, deciding when, whom, and what style to communicate in, only when it is beneficial~\cite{Tokadli2019InteractionPF}. Prior approaches that autonomously learn communication protocols have attempted to emulate similar behavior to the communication protocols used in human teams but have fallen short on assessing the heterogeneity within teams. Typical communication patterns across humans widely differ based on the responsibility or role the human assumes~\cite{Taylor2007TheEO, seraj2021hierarchical}.

Robots exhibit a similar heterogeneity based on their role within a team. We define a heterogeneous robot team as a group of cooperative agents that are capable of performing different tasks and may have access to different sensory information from the environment in which they collaborate to accomplish a shared objective. 
We categorize agents with similar action and observation spaces in the same \textit{class}. \textcolor{black}{In such a heterogeneous setting, communicating is not straightforward as agents do not speak the same ``language''; we consider scenarios in which agents have different action-spaces and observation inputs from the environment (i.e., due to different sensors) or may not even have access to any observation input (i.e., lack of sensors, broken or low-quality sensors).} 
\textcolor{black}{The dependency generated via sensor-lax or sensor-void agents on agents with strong sensing capabilities makes communication protocols for cooperation a requirement rather than an additional modeling technique for improving the performance of multi-agent coordination~\cite{seraj2021hierarchical, seraj2020coordinated, meneghetti2020towards, seraj2020coordinatedslides}.}

While the use of communication in MARL has become highly prevalent~\cite{das2018tarmac,sukhbaatar2016learning,liu2019multi,Jiang2018LearningAC,Zhang2013CoordinatingMR}, most prior framework fail to function in the presence of \emph{composite teams}. We define a composite team as a group of heterogeneous agents that perform different tasks according to their respective capabilities while their tasks are co-dependent on accomplishing an overarching mission. Examples of such composite teams include the service-agent transport problem~\cite{bays2015solution} and perception-action teams~\cite{akyildiz2004wireless, seraj2020firecommander}. In the latter example, perception agents (with only sensing capabilities) and action agents (with additional actuation capabilities) must collaborate to accomplish a task~\cite{akyildiz2004wireless, seraj2020firecommander, akkaya2008distributed, karimzadeh2017distributed}. As such, agents in a composite team can inherently have different state and action spaces and yet, must still communicate essential information with each other. None of the existing prior multi-agent communication learning frameworks are designed to explicitly model such heterogeneity in a robot team. Without a proper strategy to handle heterogeneous communications, agents of different classes may not be able to differentiate the heterogeneity in the globally sent and received massages and extract valuable information for decision making. Therefore, communication may become unhelpful, and could even deteriorate the MARL performance~\cite{zhang2019efficient}.

Inspired by \textcolor{black}{heterogeneous communication patterns across humans}, we design an end-to-end communication learning model based on Heterogeneous Graph Attention Networks (HetGAT) and build a communication channel to account for the heterogeneity of agents by ``translating'' the \textcolor{black}{inter-class} messages into a shared language. We integrate HetGAT layers to build our multi-agent heterogeneous communication policy network, HetNet, to enable multi-agent communication across agents with different roles and capabilities. As such, HetNet resolves the \emph{language barrier} across heterogeneous agents and allows for a stylized encoding and decoding of state information. 
\textcolor{black}{
Furthermore, we design our communication protocol, HetNet, with the rationale of Centralized Training and Distributed Execution (CTDE)~\cite{foerster2018counterfactual, seraj2019safe} to learn efficient, end-to-end communication models for multi-robot environments with heterogeneous agents. We propose Multi-Agent Heterogeneous Actor-Critic (MAHAC) to learn from scratch, a scalable and efficient yet effective communication protocol under Multi-Agent Heterogeneous Partially Observable (MAH-POMDP) environments. 
}

\noindent\textbf{Contributions --} The primary contributions of our work are: 
\begin{enumerate}
    \item Formulating a new problem setup, termed as Multi-Agent Heterogeneous Partially Observable Markov Decision Process (MAH-POMDP) to model a generic MARL framework for learning heterogeneous communication protocols for different classes of agents (of different types and capabilities). \textcolor{black}{We develop a novel on-policy Actor-Critic (AC) algorithm, Multi-Agent Heterogeneous Actor-Critic (MAHAC), to learn \textit{class-wise} policies for agents to enable collaborative decision-making and multi-agent cooperation.}
    
    \item Designing a limited-length communication channel to generate real-valued messages embedded in Graph Neural Network (GNN) architectures, resulting in an efficient and scalable communication model. 
    
    \item Proposing and evaluating several MAHAC architectures to study and investigate the utility of \textit{fully-centralized} critic (i.e., one critic signal for all agents of all classes), \textit{per-class} critics (i.e., one critic signal per class of agents) or \textit{per-agent} critics (i.e., individual critic signals for each agent) to learn class-wise policies for enabling heterogeneous communication and coordination.
    
    \item Evaluating our framework in a complex multi-agent heterogeneous environment (referred to as the Predator-Capture domain) in which, a composite team of agents (predators) must communicate and collaborate to capture hidden targets (preys). In this domain, collaboration is \textit{required} to satisfy the mission. 
\end{enumerate}


\section{Related Work}
\label{sec:RelatedWork}
\noindent There has been large success in generating high-performing cooperative teams using MARL in challenging problems such as game playing~\cite{berner2019dota,Samvelyan2019TheSM,peng2017multiagent} and robotics~\cite{Sun2020ScalingUM}. Recently, the use of communication in MARL has become highly prevalent as agents can share important information to greatly improve team performance \cite{das2018tarmac,sukhbaatar2016learning,liu2019multi,Jiang2018LearningAC,Zhang2013CoordinatingMR}. In this section, we discuss the body of relevant literature, including communication learning in MARL, application of GNNs in MARL and, heterogeneous multi-agent systems.

\textbf{MARL with Communication} -- Communication has been shown to further enhance the collective intelligence of learning agents in cooperative MARL problems~\cite{singh2018learning}. In recent years, several studies have been concerned with the problem of learning communication protocols and languages to use among agents. Here, we explore those bearing the closest resemblance to our work. Differentiable Inter-Agent Learning (DIAL)~\cite{foerster2016learning} and CommNet~\cite{sukhbaatar2016learning}, two of the earliest works in learning communication in MARL, displayed the capability to learn a discrete and continuous communication vectors, respectively. While DIAL considers the limited-bandwidth problem for communication, neither these approaches are readily applicable to composite teams or capable of performing attentional communication. TarMAC~\cite{das2018tarmac} on the other hand achieves targeted communication through an attention mechanism which greatly improves performance compared to prior work. Nevertheless, TarMAC requires high-bandwidth message passing channels and its architecture is reported to perform poorly in capturing the topology of interaction~\cite{liu2019multi}. SchedNet~\cite{kim2019learning} is another more recent work that explicitly addresses the bandwidth-related concerns. However, in SchedNet agents learn how to schedule themselves for accessing the communication channel, rather than learning the communication protocols from scratch. In our approach, we explicitly address the heterogeneous communication problem where agents learn diverse communication protocols and languages to use among themselves for cooperation. Our model enables agents to perform attentional communication and sending limited-length messages through class-specific communication channels, addressing the limited-bandwidth issues.

\textbf{MARL with Graph Neural Networks} -- \textcolor{black}{Graph Neural Networks (GNNs) are a class of deep neural networks that learn from unstructured data by representing objects as nodes and relations as edges and aggregating information from nearby nodes \cite{wang2020learningRAL,Wu2020ACS,Jiang2020GraphCR}.} MARL has utilized GNNs to model a communication structure among agents. Deep graph network~\cite{Jiang2020GraphCR} represents dynamic multi-agent interaction as a graph convolution to learn cooperative behaviors. This seminal work in utilizing graphs for MARL demonstrates that utilizing a graph based representation substantially improves performance in multi-agent cooperation. In \cite{sheng2020learning}, an effective communication topology is proposed by using hierarchical graph neural network to propagate messages among groups and agents. G2ANet~\cite{liu2019multi} proposed a game abstraction method combining a hard and a soft-attention mechanism to dynamically learn interactions between agents. More recently, MAGIC~\cite{niu2021multi} introduced as a scalable, attentional communication model to determine when to communicate and how to process messages, through graph attention networks and learning a scheduler. While prior work in using GNNs in MARL have successfully modeled multi-agent interactions, the mentioned frameworks are not designed to address heterogeneous robot teams. Our proposed framework on the other hand, learns an efficient shared language across agents with different action and observation spaces.

\textbf{Heterogeneity in Multi-agent Systems} -- Multi-agent communication is fundamental in composite teams, especially in the case where some agents are vision-limited. In~\cite{multi_agent_communication_hetero}, several types of heterogeneity induced by agents of different capabilities are presented and discussed. Heterogeneity of agents in a team, makes it excessively challenging to hand-design communication protocols~\cite{multi_agent_communication_hetero}. In~\cite{xiong}, a control scheme is designed for a heterogeneous multi-agent system by modeling the interaction as a leader-follower system. While this approach is successfully applied to a UAV-UGV team, the control scheme is hand-designed and requires a fully connected communication structure. More recently, HMAGQ-Net~\cite{meneghetti2020towards} utilized GNNs and Deep Deterministic Q-network (DDQN) to enable communication and cooperation among heterogeneous teams including agents with different state and action spaces. In HetNet, we build our model based on actor-critic framework and consider all state-space, action-space and observation-space heterogeneities. Moreover, we propose and assess several MAHAC architectures, investigating the utility of learning \textit{fully-centralized}, \textit{per-class} or \textit{per-agent} critics. Our framework, HetNet, can be applied to \textcolor{black}{other} composite teams and simultaneously learns a communication strategy and policy. 


\section{Problem Statement and Setup}
\label{sec:ProblemStatement&Setup}
\subsection{Problem Formulation}
\label{subsec:ProblemFormulation}
\noindent Founding on a standard Partially Observable MDP (POMDP), we formulate a new problem setup termed as Multi-Agent Heterogeneous POMDP (MAH-POMDP), which can be represented by a 9-tuple $\langle \mathcal{C}, \mathcal{N}, \{\mathcal{S}^i\}_{i\in\mathcal{C}}, \{\mathcal{A}^i\}_{i\in\mathcal{C}}, \{\Omega^i\}_{i\in\mathcal{C}}, \{\mathcal{O}^i\}_{i\in\mathcal{C}}, r, \mathcal{T}, \gamma \rangle$. $\mathcal{C}$ is set of all available agent classes in the composite robot team and the index $i\in\mathcal{C}$ shows which class does an agent belong to. $\mathcal{N}=\sum_{\langle i\in\mathcal{C}\rangle}\mathcal{N}_i$ is the total number of interacting agents in the environment in which $\mathcal{N}_i$ represents the number of agents in each class. State space $\{\mathcal{S}^i\}_{i\in\mathcal{C}}$ is a discrete set of joint states and can be factored as $\{\mathcal{S}^i\}_{i\in\mathcal{C}}=\bar{\mathcal{S}}\times\mathcal{S}^e$ where $\mathcal{S}^e$ denotes the environmental states and $\bar{\mathcal{S}}=\times_i\mathcal{S}^{(i)}$ is the joint state space of all agent classes. We note that, an agent's state-space content (e.g., the number of state variables) is determined by its class. As such, for each $\mathcal{S}^{(i)}$ we have $\mathcal{S}^{(i)}=\left[s_t^{i_j}\right]\in\bar{\mathcal{S}}$ where $s_t^{i_j}$ represents states of agent $j$ of the $i$-th class, at time $t$. Action space, $\{\mathcal{A}\}_{i\in\mathcal{C}}$, is a discrete set of joint actions, which can be factored as $\{\mathcal{A}\}_{i\in\mathcal{C}}=\times_i\mathcal{A}^{(i)}$, where $\mathcal{A}^{(i)}$ is the action space for agents of class $i$. For each $\mathcal{A}^{(i)}$ we have $\mathcal{A}^{(i)}=\left[a_t^{i_j}\right]\in\mathcal{A}$, forming the vector of joint actions. $\{\Omega\}_{i\in\mathcal{C}}$ is the class-specific observation-space, and $\gamma\in[0, 1)$ is the temporal discount factor for each unit of time. We emphasise that in our MAH-POMDP, the class of an agent determines the content of its state, action and observation spaces, such as the number of or the type of variables in these spaces.  

At each time-step, $t$, and depending on if the environment observation input is enabled for class $i$, each agent, $j$, of the $i$-th class can receive a partial observation $o_t^{i_j}\in\Omega$ according to some class-specific observation function $\{\mathcal{O}^i\}_{i\in\mathcal{C}}: o^{i_j}_t\sim\mathcal{O}^i(\cdot|\bar{s})$. If the environment observation is not available for agents of class $i$, agents in the respective class do not have access to the environmental state space, $\mathcal{S}^e$, and thus, will not receive any input from the environment. Regardless of receiving an observation from the environment or not, at each time-step, $t$, each agent, $j$, of the $i$-th class takes an action, $a_t^{i_j}$, forming a joint action vector $\bar{a}=\left(a_t^{1_1}, a_t^{1_2}, \cdots, a_t^{i_1}, \cdots, a_t^{i_j}\right)\in\mathcal{A}$. When agents take the joint action $\bar{a}$, in the joint state $\bar{s}$ and depending on the next joint-state, they receive an immediate reward, $ r(\bar{s}, \bar{a})\in\mathbb{R} $, shared by all agents, regardless of their classes. Such shared reward, encourages collaboration and teaming behaviour among agents~\cite{kim2019learning}. This step leads to changing the joint states to $\bar{s}'\in\mathcal{S}$ according to the state transition probability density function $\mathcal{T}\left(\bar{s}'|\bar{s}, \bar{a}\right)$. Our objective is to learn an optimal policy $\pi^*(\bar{s}): \mathcal{S} \rightarrow \mathcal{A}$, that solves the MAH-POMDP by maximizing the total expected, discounted reward accumulated by agents over an infinite horizon, as in Equation~\ref{eq:GeneralObjective}.
\begin{equation}
	\label{eq:GeneralObjective}
	\pi^*(\bar{s}) = \argmax_{\pi(\bar{s})\in \Pi}\mathbb{E}_{\pi(\bar{s})}\left[\sum_{k=0}^{\infty}\gamma^kr_{t+k}|\pi(\bar{s})\right]
\end{equation}

\subsection{Actor-Critic (AC)}
\label{subsec:AC}
\noindent In Actor-Critic (AC) methods, the policy of an agent $\pi^j$ is parametrized by $\theta$, represented by $\pi_\theta^j(s)$. In AC, an agent's goal is to maximize the \textcolor{black}{total expected return} by applying gradient ascent and directly adjusting the parameters of its policy, $\pi_\theta^j(s)$, through an \textit{actor} network. The actor, updates the policy distribution in the direction suggested by a \textit{critic}, which in some cases estimates the action-value function $Q^w(s, a)$~\cite{tesauro1995temporal}. For the single-agent case and by the policy gradient theorem~\cite{sutton2018reinforcement}, \textcolor{black}{the expected reward maximization (the AC objective), $J(\theta)$, is maximized via Equation~\ref{eq:ACObj}}, in which $a_t^j$, $o_t^j$ and $s_t^j$ are the action, observation and state of agent $j$ respectively, and $s_t^e$ shows the environment state at time $t$.
\begin{equation}
    \label{eq:ACObj}
    \nabla_\theta J(\theta) = \mathbb{E}_{\pi_\theta^j}\left[\nabla_\theta \log\pi_\theta^j(a_t^j|o_t^j)Q^w(s_t^j, s_t^e, a_t^j)\right]
\end{equation}

We note that in Equation~\ref{eq:ACObj}, the observation input, $o_t^{j}$, may not be available for an agent depending on the agent's class. For instance, in a \textit{perception-action} composite robot team\footnote{Perception-Action composite teams are composed of two classes of agents: (1) \textit{Perception} agents that can only sense the environment and, (2) \textit{Action} agents that can take specific action but cannot sense. See~\cite{seraj2020firecommander} for examples of such heterogeneous teams and their applicability in real-world problem.}, the action agents are not capable of sensing the environment and thus the observation input will not appear \textcolor{black}{within the action agent's policy network}. Such an agent takes actions only based on the critic signal and the communication messages, $m_t^k$, received from its neighboring agents. Due to such dependency and heterogeneity in our problem setting, \textcolor{black}{we design an actor-critic algorithm for our} multi-agent heterogeneous scenario. We refer to this as Multi-agent Heterogeneous Actor-Critic (MAHAC). MAHAC, which is introduced in Equation \ref{eq:MAHACObj} (Section~\ref{sec:TrainingandExecution}). MAHAC enables learning coordination policies (i.e., centralized, class-wise, or agent-wise) for heterogeneous agents of a composite team with different capabilities. 


\begin{figure}
	\centering
	\includegraphics[width=\columnwidth]{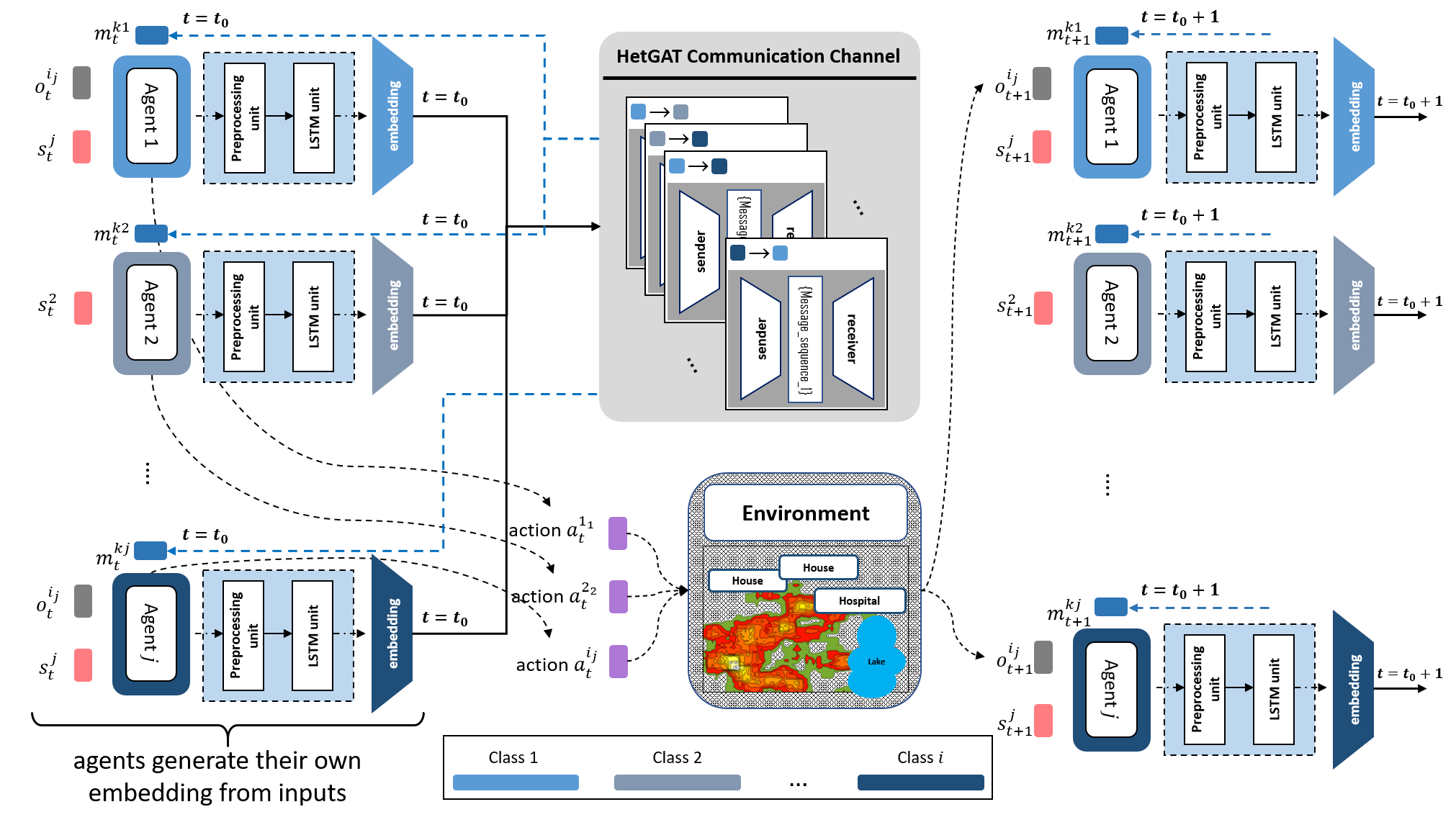}
	\caption{Overview of our multi-agent heterogeneous attentional communication architecture. At each time point $t=t_0$, each agent $j$ of class $i$ generates a local embedding from its own inputs, by passing its input data through class-specific preprocessing (i.e., a convolutional or a fully connected network) and LSTM units. Each agent then sends the embedding to a class-specific communication channel to send and receives a message, $m_t^{kj}$, from its local neighbors $k$. The message information is leveraged by the receiving agent to compute the action probabilities as its policy output.}
	\label{fig:FullMessagePassingSystem}
	\vspace*{-0.5cm}
\end{figure}

\subsection{Graph Neural Networks}
\label{subsec:GNN}
\noindent GNNs capture the structural dependency among nodes of a graph via message-passing between the nodes, in which each node aggregates feature vectors of its neighbors to compute a new feature vector. The commonly used feature update procedure via graph convolution operator is shown in Equation~\ref{eq:gnn}, where $\bar{h}_j^\prime$ is the updated feature vector for node $j$, $\sigma(.)$ is the activation function and, $\omega$ represents the learnable weights. 
\begin{equation}
    \bar{h}_j^\prime = \sigma\left(\sum_{k\in N(j)} \frac{1}{c_{jk}}\omega\bar{h}_k\right)
    \label{eq:gnn}
\end{equation}In Equation \ref{eq:gnn}, $k\in N(j)$ includes the immediate neighbors of node $j$ where $k$ is the index of neighbor, and $c_{jk}$ is the normalization term which depends on the graph structure. A common choice of $c_{jk}$ is $\sqrt{|N(j)N(k)|}$. After $L$ layers of aggregation, a node $i$'s representation captures the structural information within the nodes that are reachable from $i$ in $L$ hops or fewer. However, the fact that $c_{jk}$ is structure-dependent can hurt generalizability of GNN on varying graph sizes. Accordingly, a direct improvement over Equation~\ref{eq:gnn} is to replace $c_{jk}$ with attention coefficients, $\alpha_{jk}$, computed via Equation~\ref{eq:attention}. In Equation~\ref{eq:attention}, $\bar{a}$ is the learnable weight, $\mathbin\Vert$ represents concatenation, and $\sigma^\prime(.)$ is the LeakyReLU nonlinearity. The Softmax function is used to normalize the coefficients across all neighbors $k$, enabling feature dependent and structure free normalization~\cite{velickovic2018graph}.
\begin{equation}
\label{eq:attention}
\alpha_{jk} = \mathrm{softmax}_k\left(\sigma^\prime\left(\bar{a}^T\left[\omega\bar{h}_j \mathbin\Vert\omega\bar{h}_k\right]\right)\right)
\end{equation}

Heterogeneous GNNs, which directly operate on heterogeneous graphs containing different types of nodes and edges, can learn per-edge-type message passing and per-node-type feature reduction mechanisms. Compared to homogeneous GNNs, heterogeneous GNNs have shown good interpretability and model expressiveness~\cite{wang2019heterogeneous, wang2020heterogeneousRSS}. However, such a model has never been applied to MARL problems, and to the best of authors' knowledge, this is the first attempt to leverage heterogeneous GNNs for end-to-end learning of diverse communication protocols in a heterogeneous multi-robot team.

\textcolor{black}{In a heterogeneous setting, where agents have different action-spaces, communicating is not straightforward, as agents do not speak the same ``language.'' In the following sections, we design a heterogeneous communication learning structure based on Graph Attention Networks (GAT) and build an encoder-decoder communication channel to account for the heterogeneity of our agents by ``translating'' the inter-agent messages into a shared language.}


\section{Communication Problem and Overview}
\label{sec:CommunicationProblemandOverview}
\noindent \textcolor{black}{Common multi-agent coordination problems} are centered around two ideas: (1) fostering direct communication among interacting agents~\cite{foerster2016learning, sukhbaatar2016learning, kim2019learning} or, (2) coordinating agents' actions for cooperation without direct communication~\cite{foerster2017stabilising, leibo2017multi, palmer2017lenient}. In this work, we are concerned with the former. We consider MARL problems wherein multiple agents interact in a single environment to accomplish a task which is of a cooperative nature. We are particularly interested in scenarios in which the agents are heterogeneous in their capabilities, meaning agents can have different state, action and observation spaces, forming a composite team. To \textcolor{black}{collaborate effectively,} agents must share messages that express their observations and experiences under a CTDE paradigm~\cite{foerster2018counterfactual, kim2019learning}.

An overview of our multi-agent heterogeneous attentional communication architecture is shown in Figure~\ref{fig:FullMessagePassingSystem}. In learning an end-to-end communication model, we take a series of problems and constraints into consideration: (1) heterogeneous communication, where agents of different classes have different action and observation spaces, resulting in different interpretations of sent-received messages.
(2) Attentional and scalable communication protocols such that agents incorporate attention coefficients depending on the agent/class they are communicating with for coordinating with teammates in any arbitrary team sizes.


\section{HetNet for Learning Multi-agent Heterogeneous Communication}
\label{sec:CommunicationChannel}
\noindent \textcolor{black}{GNNs previously used in MARL} operate on homogeneous graphs to learn a universal feature update and communication scheme for all agents, which fails to \textcolor{black}{explicitly model} the heterogeneity among agents \textcolor{black}{in our case}. We instead cast the MARL problem into a heterogeneous graph structure, and propose a novel heterogeneous graph attention network capable of learning diverse communication strategies based on agent classes. In this section, we first describe how to construct the heterogeneous graph given a problem state. Then, we present the building block layer, which we refer to as heterogeneous graph attention (HetGAT) layer, used to assemble our heterogeneous policy network (HetNet) of arbitrary depth (Section~\ref{subsec:HetPolicyNetwork}).

\subsection{Heterogeneous Graphs for Composite Multi-Agent Teams}
\label{subsec:HeterogeneousGraphForCompositeMulti-AgentTeams}
\noindent Compared to homogeneous graphs, a heterogeneous graph can have nodes and edges of different types. Such different types of nodes and edges tend to have different types of attributes that are designed to capture the characteristics of each node and edge. This advantage greatly increases a graph's expressivity and enables straightforward modeling of complicated multi-agent teams, such as our composite robot teams~\cite{seraj2020firecommander}.

Given our \textcolor{black}{MAH-POMDP formulation discussed in Section~\ref{subsec:ProblemFormulation}, we directly model each agent class in $\mathcal{C}$ as a unique node type. This allows agents to have different types of state-space content, $\mathcal{S}^{(i)}$, as input features according to their classes, $i\in\mathcal{C}$, as well as enabling different types of action spaces, $\mathcal{A}^{(i)}$.} For instance, in our perception-action composite team example in Section~\ref{subsec:AC}, the input features of perception agents contain their sensor input image and the state vector, while action agents input only contains their state vector. Communication channel between agents is modeled as directed edges connecting the corresponding agent nodes. When two agents move to a close proximity of each other such that they fall within the communication range, bidirectional edges are added to allow message passing between them. We use different edge types to model different combinations of the sender agent's class and the receiver agent's class to allow for learning heterogeneous communication protocols.

\textcolor{black}{When training the policy network constructed through stacking several HetGAT layers, we leverage our extended MAHAC (introduced in Section~\ref{subsec:MAHAC}) to learn class-wise agent policies. To enable Centralized Training and Distributed Execution,} we add a State Summary Node (SSN) into the graph and develop novel architectures to learn a centralized critic network, per-class critics or, per-agent critic signals. We introduce and investigate these architectures in Section~\ref{subsec:critic_version}. Depending on the selected HetNet critic architecture, the SSN forms a one-way connection to the agent nodes (directed from an agent to the SSN) to receive messages from them during training phase. The initial input features to the SSN are hyper-parameter information of the environment, such as the total number of agents, $\mathcal{N}$, world size, \textcolor{black}{current time step}, etc. The SSN's learned embeddings are used as input of a critic network consisting of one fully-connected layer for state-dependant value estimation. We note that, since there are no edges pointing from the SSN to any agent nodes, during the execution phase, the SSN can be safely removed without affecting an agent's own policy output, which complies with our underlying CTDE paradigm.

\subsection{HetGAT Layer with Heterogeneous Communication Channel}
\label{subsec:HetGATwithDigitizedCommunicationChannel}
\begin{figure}[t!]
	\centering
	\includegraphics[width=\linewidth]{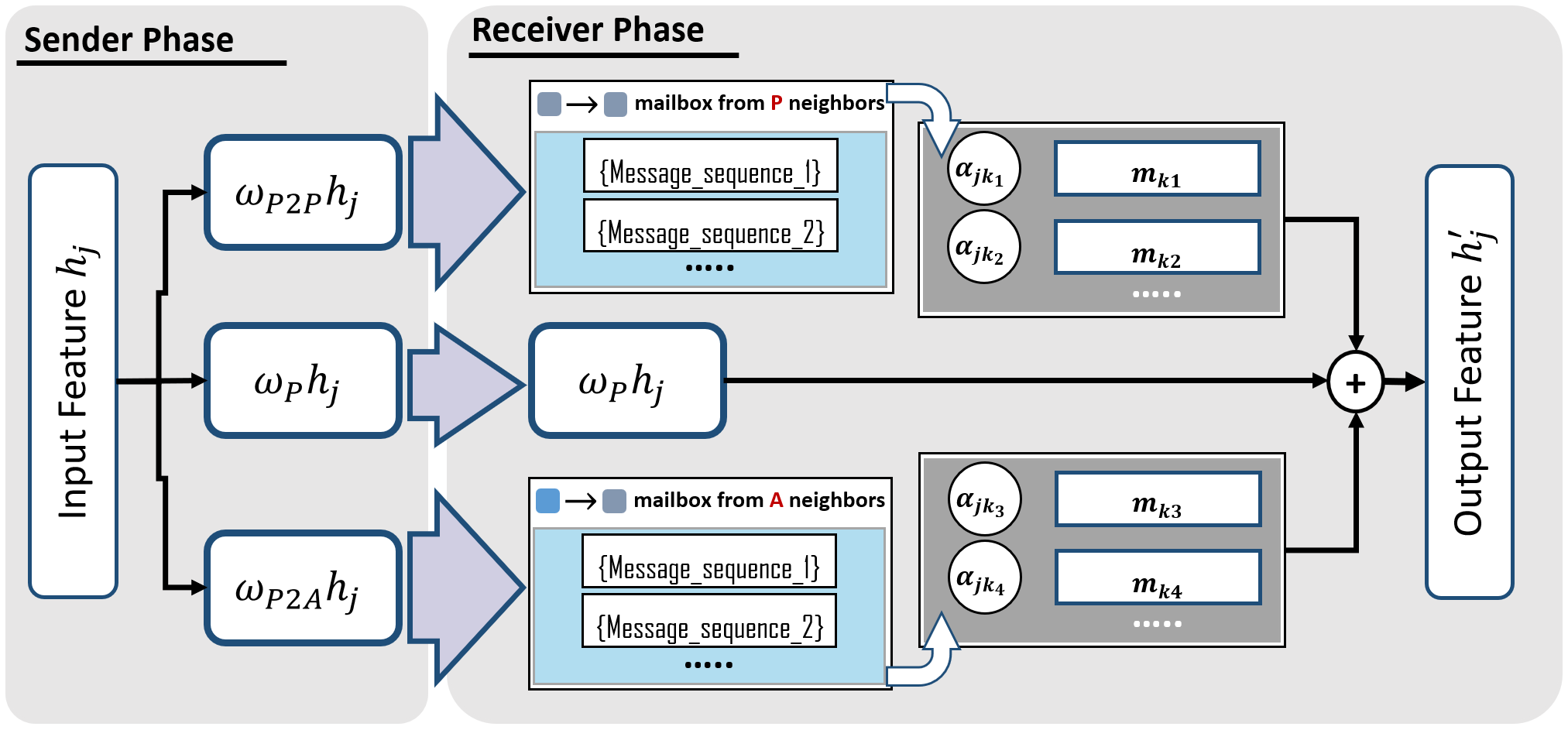}
	\caption{The sender and receiver phases of the feature update process in a HetGAT layer, using $\mathcal{C}=\{P, A\}$ as a composite team and the $P$ agent node in this team as an example.}
	\label{fig:graphcompute}
	\vspace*{-0.5cm}
\end{figure}

\noindent The feature update process in a HetGAT layer is conducted in two steps: per-edge-type message passing followed by per-node-type feature reduction. When modeling multi-agent teams, we reformulate the computation process into two phases: a sender phase and a receiver phase. For notation simplicity and without loosing generality, we demonstrate these phases through an example of a composite robot team with two classes of agents, $\mathcal{C}=\{P, A\}$ (e.g., see~\cite{seraj2020firecommander}). Figure~\ref{fig:graphcompute} shows the computation flow during the sender and receiver phases with P agent node as an example.

During the sender phase, the agent of class $P\in\mathcal{C}$ (referred to as P agent), indexed by $j$, processes its input feature, $h_j$, using a class-specific weight matrix, $\omega_{P} \in \mathbb{R}^{d^\prime \times d}$, where $d$ is the input feature dimension, and $d^\prime$ is the output feature dimension. Meanwhile, each type of communication channel uses a \textcolor{black}{distinct weight matrix, $\omega_{edgeType} \in \mathbb{R}^{d'' \times d}$,} to process $h_j$ (e.g., $\omega^{PtoA}$, denoting communication edge type from $P$ agents to $A$ agents), and sends the computation results to the mailbox of the destination node, $N_{dst}$. \textcolor{black}{Here,} $d''$ is the output feature dimension of $N_{dst}$. 

During the receiver phase, for each type of the communication edge that an agent is connected to, the HetGAT layer computes per-edge-type aggregation result by weighing received messages, stored in its mailbox, along the same edge type with normalized attention coefficients, $\alpha^{edgeType}$. The aggregation results are then merged with the agent's own transformed embedding, $\omega_P h_j$, to compute the output feature. As a results, the feature update formula for P agents can be shown as in Equation~\ref{eq:Pnode} where $N_P(j)$ and $N_A(j)$ include agent $j$'s neighbors that are of class $P$ and $A$, respectively.
\par\nobreak{\parskip0pt \normalsize \noindent
\begin{equation}
    \label{eq:Pnode}
    \text{Class}~P~\bar{h}_j^\prime = \sigma\Big(\omega_P \bar{h}_j +\sum_{k \in {N_P(j)}}\alpha_{jk}^{PtoP} \omega_{PtoP} \bar{h}_k + \sum_{l\in{N_{A}(j)}}\alpha_{jl}^{AtoP}  \omega_{AtoP} \bar{h}_l\Big)
\end{equation}}Note that, when computing attention coefficients in a heterogeneous graph, we adapt Equation~\ref{eq:attention} into Equation~\ref{eq:attNew} to account for heterogeneous communication channels.
\begin{equation}
\label{eq:attNew}
    \alpha_{jk}^{edgeType} = \mathrm{softmax}_k\left(\sigma^\prime\left(\bar{a}^T\left[\omega_P\bar{h}_j \mathbin\Vert \omega_{edgeType} \bar{h}_k \right]\right)\right)
\end{equation}A similar computation process applies for $A$ agents, with the update formula shown in Equation~\ref{eq:Anode}.
\par\nobreak{\parskip0pt \normalsize \noindent
\begin{equation}
    \label{eq:Anode}
    \text{Class}~ A~\bar{h}_j^\prime = \sigma\Big(\omega_A \bar{h}_j +\sum_{k\in{N_P(j)}} \alpha_{jk}^{PtoA} \omega_{PtoA}\bar{h}_k + \sum_{l \in {N_A(j)}} \alpha_{jl}^{AtoA} \omega_{AtoA} \bar{h}_l \Big) 
\end{equation}}As discussed in Section~\ref{subsec:HeterogeneousGraphForCompositeMulti-AgentTeams}, we add an SSN to the graph during centralized training with a state-dependent critic network. The feature update formula of SSN is shown in Equation~\ref{eq:Statenode}. Here, feature vectors from all agents are passed to SSN after being processed with edge-specific weights, $\omega_{edgeType}$. For SSN, the attention coefficients are computed in a similar manner as in Equation~\ref{eq:attNew}.
\par\nobreak{\parskip0pt \normalsize \noindent
\begin{equation}
    \label{eq:Statenode}
    \text{SSN}~\bar{h}_s^\prime = \sigma \Big(\omega_S \bar{h}_s +\sum_{k \in P} \alpha_{sk}^{PtoS} \omega_{PtoS} \bar{h}_k + \sum_{l \in A} \alpha_{sl}^{AtoS} \omega_{AtoS} \bar{h}_l \Big) 
 \end{equation}}We reiterate that equations~\ref{eq:Pnode}-\ref{eq:Statenode} \textcolor{black}{are generally applicable to any heterogeneous robot team with arbitrary number of agent classes and are not restricted to the $P$ and $A$ agents example.}
 
Finally, to stabilize the learning process, we utilize the multi-head extension of the attention mechanism, proposed in~\cite{velickovic2018graph}, adapting it to fit the heterogeneous teams. We use $K$ independent HetGAT (sub-)layers to compute node features in parallel, and then merge the results as the multi-head output by concatenation operation for each multi-head layer in HetNet, except for the last layer which employs averaging. As a result, each type of communication channel is split into $K$ independent sub-channels.

\subsection{Heterogeneous Policy Network (HetNet)}
\label{subsec:HetPolicyNetwork}
\noindent At each timestep, a HetGAT layer corresponds to one round of message exchange between neighboring agents and feature update within each agent. By stacking several HetGAT layers, we construct Heterogeneous Policy Network (HetNet) model that utilizes multi-round communication to extract high-level embeddings of each agent for decision-making. For the last HetGAT layer in HetNet, we set each agent's output feature dimension the same size as its action-space, specific to its class. Then, for each agent node, we add a Softmax layer on top of its output to obtain a probability distribution that can be used for action sampling, resulting in class-wise stochastic policies. By doing so, the computation process of each agent's policy remains local for distributed execution, and the SSN is no longer needed during execution/testing.

\textbf{Feature Preprocessing} -- In HetNet, we utilize separate modules to preprocess an agent's state vector, $s_t$, and observation, $o_t$ (if applicable depending on the class) into $o^\prime_t$ and $s^\prime_t$, respectively, before using them as input node features. Each preprocecssing module contains one fully-connected layer followed by one Long Short-Term Memory (LSTM) layer to enable reasoning about temporal information. Note that in the two-class perception-action composite team example, the input feature of perception agents ($P$ agent) node is the concatenation of $o^\prime_t$ and $s^\prime_t$, while for action agent ($A$ agent) nodes the input feature is $s^\prime_t$.


\section{Training and Execution}
\label{sec:TrainingandExecution}
\subsection{Multi-agent Heterogeneous Actor-Critic (MAHAC)}
\label{subsec:MAHAC}
\noindent The use of heterogeneous GNNs enables us to deploy our learning framework under the CTDE paradigm. \textcolor{black}{Here, we present an Actor-Critic approach to fit our multi-agent heterogeneous scenario, Multi-agent Heterogeneous Actor-Critic (MAHAC).} Due to heterogeneity of the action-spaces, we assign one policy to each class of agents, represented in a joint policy vector $\bar{\pi}=\left(\pi^1, \pi^2, \cdots, \pi^i\right)\in\{\Pi\}^{|\mathcal{C}|}$. We parametrize each policy in $\bar{\pi}$ by its respective parameter from the joint vector $\bar{\theta}=\left(\theta^1, \theta^2, \cdots, \theta^i\right) \text{for}~ i=1, ..., |\mathcal{C}|$ and learn one policy network per each class of agents, while one centralized critic network ``criticizes'' the actions of all agents. Note that, this approach still complies with our CTDE paradigm, since the actor network is implemented on a GNN structure. The trained GNN contains one set of learnable weights per agent class, which due to the message-passing nature of GNN updates, can be distributed to individual agents in the execution phase. Accordingly, in MAHAC, the policy for each class, $\pi^i$, is updated by a variant of Equation~\ref{eq:ACObj}, shown in Equation~\ref{eq:MAHACObj}. We leverage an on-policy training paradigm for MAHAC.
\par\nobreak{\parskip0pt \normalsize \noindent
\begin{align}
    \nabla_{\theta^i} J(\theta^i) = \frac{1}{N}\sum_{j=1}^{\mathcal{N}_i}\sum_{t=1}^{T}\nabla_{\theta^i}\log\pi^i\left(\bar{a}_t^{i_j}|\bar{o}_t^{i_j}, m_t^k\right)\left(\left(\sum_{t'=t}^{T}\gamma^{t'-t}r(\bar{s}^j, \bar{a}^j)\right)-b(t)\right)
    \label{eq:MAHACObj}
\end{align}}

In Equation~\ref{eq:MAHACObj}, indices $j$ and $i$ refer to an agent and its class, respectively. $\bar{a}_t^{i_j}$ and $\bar{o}_t^{i_j}$ represent the joint actions taken and joint observations received by agents of class $t$ at time $t$. Note that, in Equation~\ref{eq:MAHACObj}, the observation input, $\bar{o}_t^{i_j}$, may not be available for an agent, depending on its class. $m_t^k$ represents the input communication message received by agents $j$ from its neighboring agent $k\in N_t(j)$, where $k$ is the neighbors index. The term $\sum_{t'=t}^{T}\gamma^{t'-t}r(\bar{s}^j, \bar{a}^j)$ calculates the total discounted future reward from current time-step until the end of the episode. Note that, the reward is shared by all agents (i.e., regardless of their class) and thus, only the superscript $j=1, \cdots, \mathcal{N}$ appears over the joint states and actions in $r(\bar{s}^j, \bar{a}^j)$. Moreover, $b(t)$ is a temporal baseline function leveraged to reduce the variance of the gradient updates in MAHAC. \textcolor{black}{We utilize the value estimation via our critic network as the baseline function.}

\subsection{Different HetNet Architectures and Critic Choices}
\label{subsec:critic_version}
\begin{figure}
    \centering
    \begin{subfigure}[t]{0.28\textwidth}
        \centering
        \includegraphics[width = \linewidth]{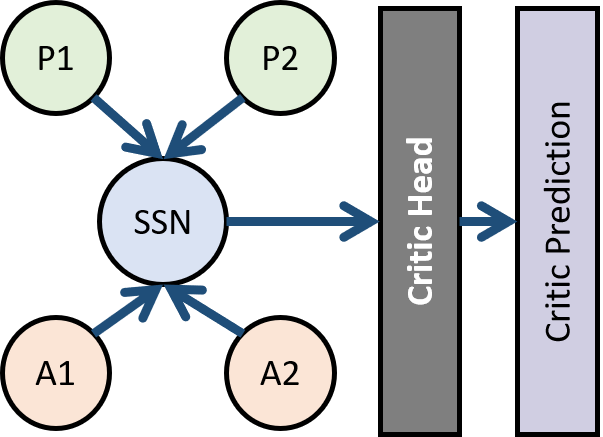}
        \caption{One centralized critic}\label{fig:D3}
    \end{subfigure}
    \begin{subfigure}[t]{0.28\textwidth}
    \centering
        \includegraphics[width = \linewidth]{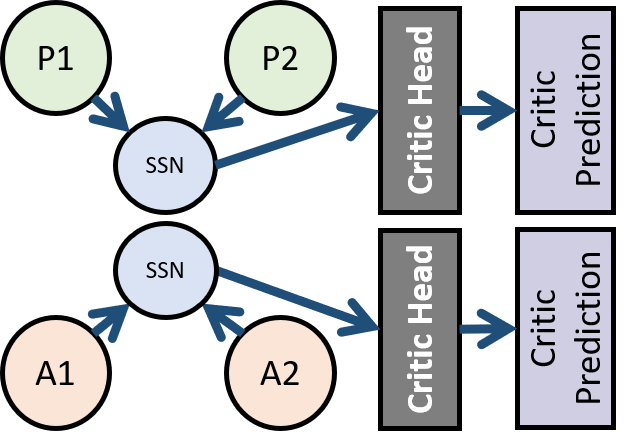}
        \caption{Per-class critic}\label{fig:D4}
        \end{subfigure}
    \begin{subfigure}[t]{0.43\textwidth}
    \centering
        \includegraphics[width = \linewidth]{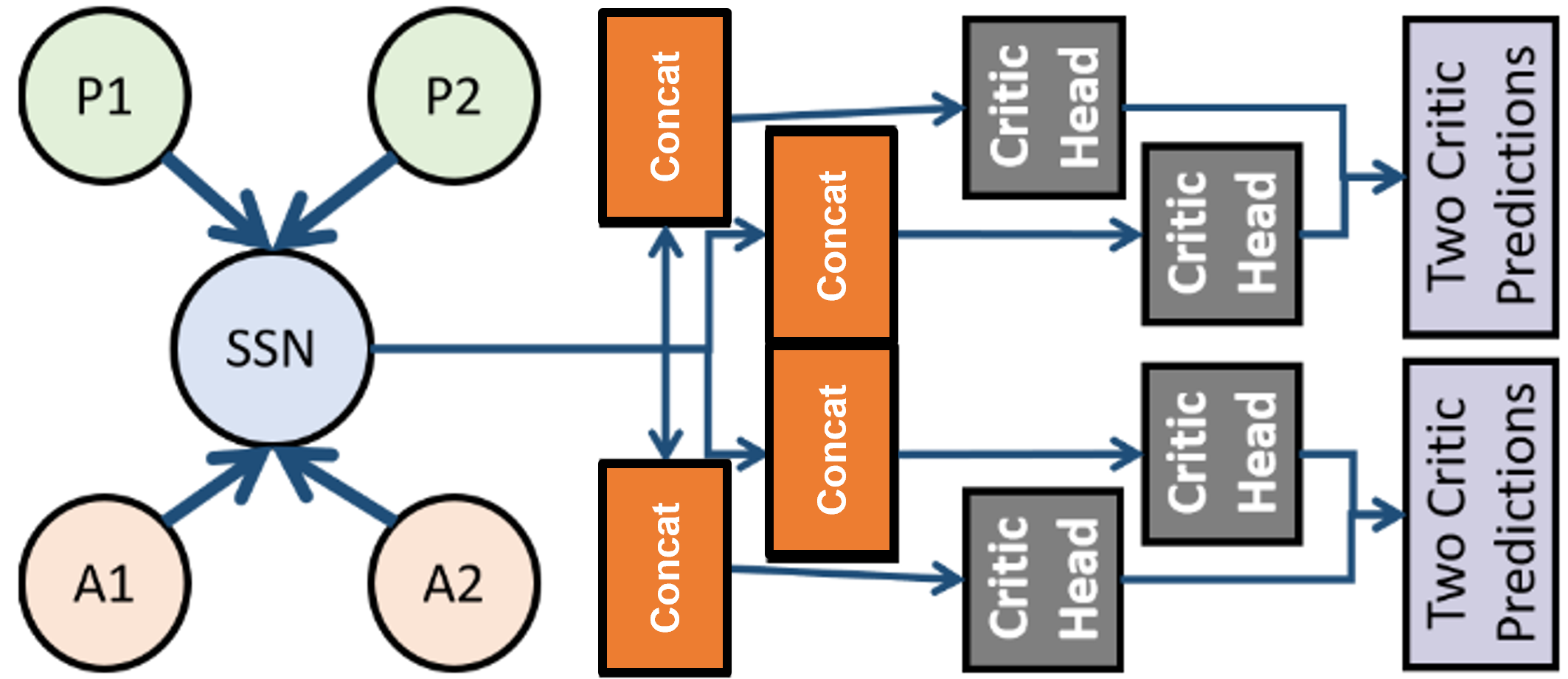}
        \caption{Per-agent critic}\label{fig:D5}
        \end{subfigure}
    \caption{Different proposed critic implementations with a perception-action composite team consisting of two $P$ agents and two $A$ agents: (a) one centralized critic, (b) per-class critics and (c) per-agent critics.}
    \label{fig:critic}
    \vspace*{-0.5cm}
\end{figure}

\noindent In this section we propose and assess several MAHAC architectures to investigate the utility and performance of: (1) \textit{fully-centralized} critic (i.e., one critic signal for all agents of all classes), (2) \textit{per-class} critics (i.e., one critic signal per class of agents) and (3) \textit{per-agent} critics (i.e., individual critic signals for each agent) to learn class-wise policies for enabling heterogeneous communication and coordination. Figure~\ref{fig:critic} illustrates different critic implementations for a perception-action composite team consisting of two $P$ agents and two $A$ agents.

Figure~\ref{fig:D3} represents our suggested fully-centralized critic implementation for HetNet in which, a fully-connected layer is stacked on top of SSN's output feature for critic prediction. The same predicted critic value is used in the policy gradient update for all agents of all classes (e.g., $P$ and $A$ in this example). The target value for training the critic output is the average returns (i.e., discounted sum of future rewards) over all agents.

For the per-class critic implementation suggested in Figure~\ref{fig:D4}, the critic head is split into one critic head per existing agent classes (e.g., $P$ critic head and $A$ critic head in this example) in order to separate the critic estimation for different types of agents. Note that this critic split is done while the critic is still estimated based on the SSN's output feature. Different types of critic predictions are used in policy gradients update of the corresponding classes of agents. During training, the target value for each class of critic output is the average returns over the same class of agents.

Figure~\ref{fig:D5} shows our suggested per-agent critic implementation for HetNet, where the critic network outputs one critic estimation for each agent. This is achieved by concatenating the SSN's output feature with each agent node's output embedding to serve as the input of class-specific critic heads. The per-agent critic estimation is used for each agent's policy update, and its target value for training is the returns of that agent.


\section{Evaluation Environments}
\label{sec:EvaluationEnvironment}
We evaluate the utility of HetNet against several baselines in both two cooperative MARL domains that require coordination and learning collaborative behaviors. \textcolor{black}{We use a common MARL homogeneous domain and a complex heterogeneous environment to test HetNet's general performance and applicability.}

\noindent\textbf{Predator-Prey (PP)~\cite{singh2018learning} --} The goal in this homogeneous environment is for $N$ predator agents with limited vision to find a stationary prey and move to its location. The agents in this domain are homogeneous in their state, observation and action spaces and thus, all agents are of the same \textit{class}. The positions of the predator agents and the prey are initialized randomly in the beginning of each episode. The state-space, for all agents, is a concatenated one-hot vector which represents an agent's own location and binary information indicating the presence of other predator agents or the prey at each timestep. All agents are able to sense/observe the environment and each agent's observation is a concatenated array of the state vectors of all grids within the agent’s field-of-view (FOV). The predator agents' action-space is of dimension five and is the same for all agents; including actions \textit{move up}, \textit{move down}, \textit{move right}, \textit{move left} and \textit{stay}. Each predator agent receives a small penalty of $-0.05$ per timestep until the prey is found. An episode of the game is labeled as successful if all predator agents find the prey and move to its location before a predefined maximum time limit. Therefore, agents are required to communicate and coordinate their actions to win the game as fast as possible. For our experiments, we chose the world-size and the number of predator agents to be $10\times10$ and $3$, respectively. We set the maximum steps for an episode (i.e., termination condition) to be $80$. 
We define a better-performing algorithm in this domain as the one that minimizes the average number of steps taken by agents to complete an episode.

\noindent\textbf{Predator-Capture-Prey (PCP) --} 
\textcolor{black}{In this environment, we have two classes of agents, \textit{predator} agents and \textit{capture} agents. Predator agents have the goal of finding the prey. The state space of predator agents is a concatenated one-hot vector which represents an agent's own location and binary information indicating the presence of other predator agents, capture agents, or the prey at each timestep. The predator agents' action-space is of dimension five and includes the actions: \textit{move up}, \textit{move down}, \textit{move right}, \textit{move left}, and \textit{stay}. The second \textit{class} of agents, the \textit{capture} agents have the goal of locating the prey \emph{and} capturing it. Capture agents differ from the predator agents in both their observation and their action spaces such that, capture agents do not receive any observation inputs from the environment (i.e., no scanning sensors). Moreover, capture agents have an additional action of \textit{capture prey} in their action-space, such that when they move to a prey's location, they need to take the \textit{capture prey} action to capture the prey sitting in the corresponding grid. As such, the goal of the game is for all predator agents to find a stationary prey and for all capture agents to find the prey and then capture it. Agents will receive a $-0.05$ per timestep reward until they accomplish their per-class objective.} Note that this domain is an explicit example of the perception-action composite teams~\cite{seraj2020firecommander}, as introduced in Section~\ref{subsec:AC}. Here, predator agents play the role of perception agents ($P$ class) since they can only sense and searching the environment for a hidden target while capture agents play the role of action agents ($A$ type), since they cannot sense the environment (i.e., null observation input) but instead, they need to ``act" by capturing the prey found by predator (perception) agents. All other settings in this domain are similar to the PP domain.


\section{Experiment}
\label{sec:Experiment}
\noindent In this section, we empirically evaluate our multi-agent heterogeneous communication learning model, HetNet, and present its performance results in both of the introduced domains and against two other state-of-the-art, end-to-end communication learning baselines. We also present an ablation study to investigate the effects of the critic structures proposed in Section~\ref{subsec:critic_version} on HetNet's performance.

\noindent\textbf{Model Details --} We implement HetNet using PyTorch \cite{NEURIPS2019_9015} and Deep Graph Library \cite{wang2019dgl}. The HetNet used in training/testing is constructed by stacking three multi-head HetGAT layers on top of the feature preprocessing modules. The first two multi-head layers use $K$ = 4 attention heads computing 16 features each (for a total of 64 features merged by concatenation). The final layer also uses $K$ = 4 attention heads but the output dimension is set to the same size as each agent's action space and is merged by averaging. We used Adam optimizer \citep{kingma2014adam} through training with a learning rate of $10^{-3}$ for all our experiments and results presented here. 

\subsection{Baseline Comparison Results}
\label{subsec:BaselineComparisonResults}
\begin{figure}
    \centering
    \begin{subfigure}[t]{0.45\textwidth}
        \centering
        \includegraphics[width = \linewidth]{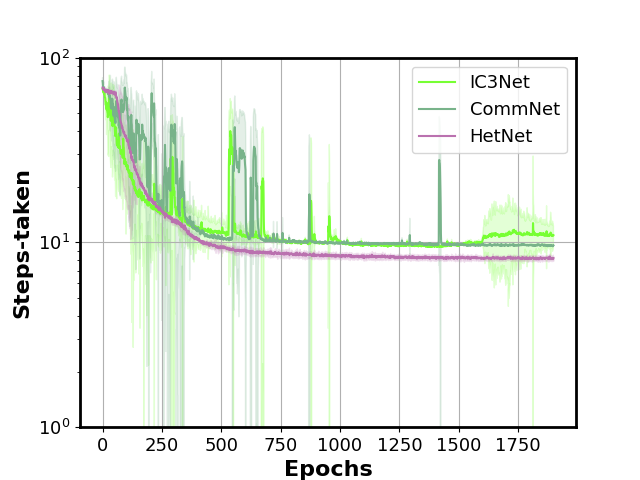}
        \caption{Homogeneous Domain (PP)}\label{fig:BaselineComp_1}
    \end{subfigure}
    \begin{subfigure}[t]{0.45\textwidth}
    \centering
        \includegraphics[width = \linewidth]{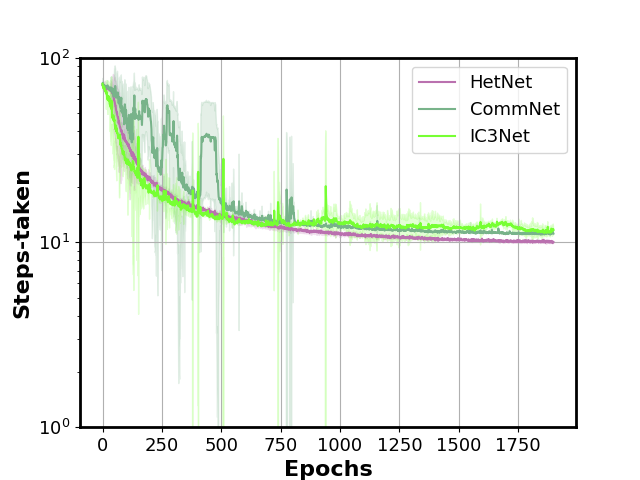}
        \caption{Heterogeneous Domain (PCP)}\label{fig:BaselineComp_2}
    \end{subfigure}
    \caption{This figures presents the learning curves during training for HetNet, CommNet~\cite{sukhbaatar2016learning} and IC3Net~\cite{singh2018learning} in the homogeneous PP (left-side) and the heterogeneous PCP (right-side) domains. Curves display the average number of steps taken on a logarithmic scale across several batches and the fewer number of steps indicates a better method. As shown, HetNet outperforms all baselines in both the homogeneous and the heterogeneous environments.}
    \label{fig:BaselineComp}
    \vspace*{-0.5cm}
\end{figure}

We benchmark our approach against two state-of-the-art, end-to-end communication learning baselines, namely the CommNet~\cite{sukhbaatar2016learning} and the IC3Net~\cite{singh2018learning}. The results are presented in Figure~\ref{fig:BaselineComp}. Figure~\ref{fig:BaselineComp} presents the learning curves during training for HetNet, CommNet~\cite{sukhbaatar2016learning} and IC3Net~\cite{singh2018learning} in the homogeneous PP (left-side) and the heterogeneous PCP (right-side) domains. Curves are showing the average number of steps taken across several batches ($\pm$ standard error) and the fewer number of steps indicates a better method. As shown, HetNet outperforms all baselines in both the homogeneous and the heterogeneous environments.

We also tested the learnt coordination policies by each of the baselines. Table~\ref{tab:BaslineComp} presents the average number of steps taken for coordination policies learnt at convergence by HetNet, CommNet~\cite{sukhbaatar2016learning} and IC3Net~\cite{singh2018learning} in the homogeneous PP (left-side) and the heterogeneous PCP (right-side) domains. Results represent the average number of steps ($\pm$ standard error) taken by the agents to win an episode of the game in 100 trials. As shown, HetNet again outperforms all baselines in both the homogeneous and the heterogeneous environments. The policies learnt by our model are more efficient and are capable of solving both PP and PCP faster than the other baselines, \textcolor{black}{using $\sim$15\% less in terms of average steps taken in PP and $\sim$10\% less in PCP}
\begin{table}[h!]
\normalsize
\caption{This table presents the average number of steps taken for coordination policies learnt at convergence by HetNet, CommNet~\cite{sukhbaatar2016learning} and IC3Net~\cite{singh2018learning} in the homogeneous PP and the heterogeneous PCP domains. Values represent the average number of steps ($\pm$ standard deviation) taken by the agents to win an episode of the game in three trials with different random-seed initialization. The bolded result in each column represents the highest-performing method. As shown, HetNet outperforms all baselines in both the homogeneous and the heterogeneous environments.}
\begin{center}
\begin{tabular}{P{55pt}P{65pt}P{65pt}P{65pt}P{65pt}}
\hline
\multirow{2}{*}{Method} & \multicolumn{2}{c}{{Homogeneous Domain (PP)}} & \multicolumn{2}{c}{{Heterogeneous Domain (PCP)}} \\
\cline{2-5} & Avg. Cumulative $\mathcal{R}$ & Avg. Steps Taken & Avg. Cumulative $\mathcal{R}$ & Avg. Steps Taken \\
\hline
\multirow{2}{*}{CommNet~\cite{sukhbaatar2016learning}} & -0.33 & 9.53 & -0.40 & 11.03 \\
& $\pm$ 0.01 & $\pm$ 0.05 & $\pm$ 0.01 &  $\pm$ 0.08 \\
\hline
\multirow{2}{*}{IC3Net~\cite{singh2018learning}} & -0.33 & 9.43 & -0.40 & 11.23 \\
&  $\pm$ 0. & $\pm$ 0.12 & $\pm$ 0.02 &  $\pm$ 0.63 \\
\hline
\multirow{2}{*}{HetNet} & \textbf{-0.30} & \textbf{8.10} & \textbf{-0.38} & \textbf{10.01} \\
& \textbf{$\pm$ 0.03}  & \textbf{$\pm$ 0.35} & \textbf{$\pm$ 0.01} & \textbf{$\pm$ 0.39} \\
\hline
\label{tab:BaslineComp}
\end{tabular}
\end{center}
\vspace{-25pt}
\end{table}

\subsection{Ablation Study: Effects of Different Critic Structures}
\label{subsec:AblationStudy}
\noindent In this section, we present an ablation study to assess different MAHAC architectures proposed in Section~\ref{subsec:critic_version}. Our goal is to investigate the utility and performance of: (1) \textit{fully-centralized} critic (i.e., one critic signal for all agents of all classes), (2) \textit{per-class} critics (i.e., one critic signal per class of agents) and (3) \textit{per-agent} critics (i.e., individual critic signals for each agent) to learn class-wise policies for enabling heterogeneous communication and coordination.

\begin{figure}
    \centering
    \begin{subfigure}[t]{0.47\textwidth}
        \centering
        \includegraphics[width = .92\linewidth]{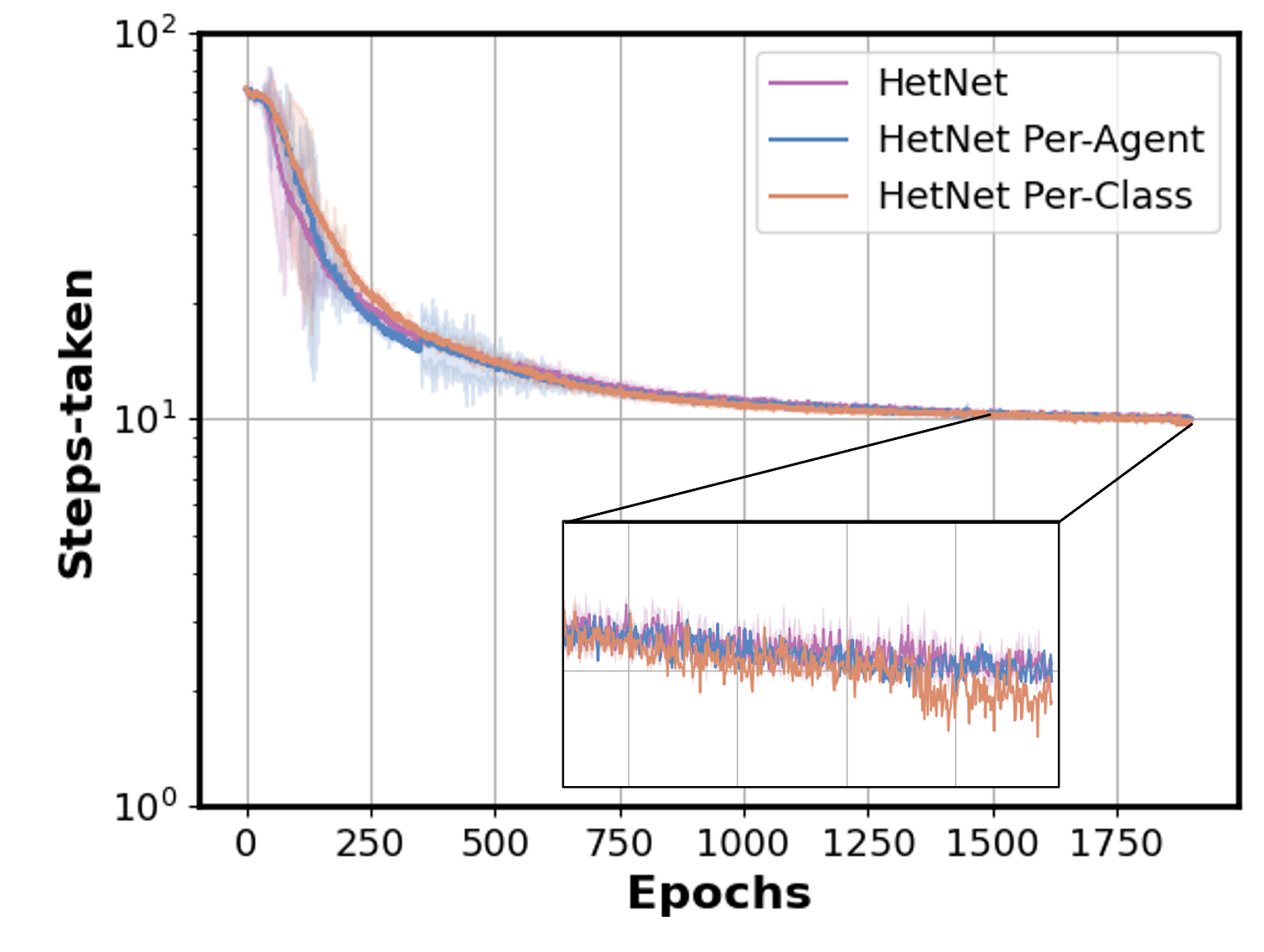}
        \caption{\textcolor{black}{Training Performance across 2000 Epochs.}}\label{fig:critic_ablation_1}
    \end{subfigure}
    \begin{subfigure}[t]{0.44\textwidth}
    \centering
        \includegraphics[width = \linewidth]{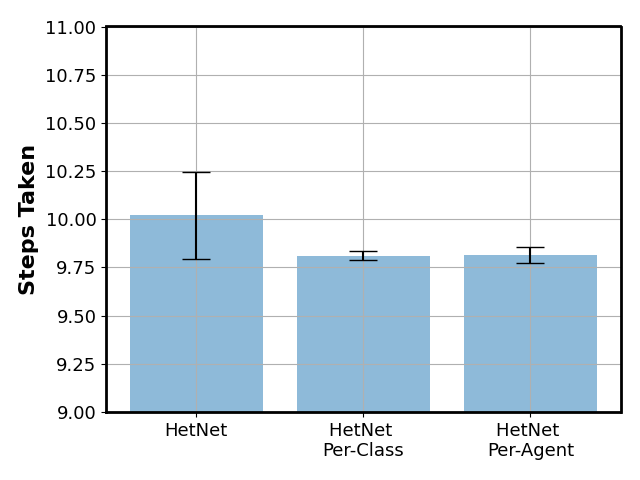}
        \caption{\textcolor{black}{Policy Performance at Convergence.}}\label{fig:critic_ablation_2}
    \end{subfigure}
    \caption{This figures presents the learning curves (left-side) during training for centralized, per-class and per-agent critic architectures (shown in Figure~\ref{fig:critic}) in the heterogeneous PCP domain. Curves are showing the average number of steps taken across several batches ($\pm$ standard error) on a logarithmic scale for each critic structure. Moreover, the test results for policies learnt by each of the critic architectures are presented in Figure~\ref{fig:critic_ablation}, right-side. This figure shows the bar-plots representing the average number of steps taken to win the game by deploying the learnt policies through centralized, per-class and per-agent critic architectures.}
    \label{fig:critic_ablation}
\end{figure}

The results of our critic structure ablation study are presented in Figure~\ref{fig:critic_ablation}. Figure~\ref{fig:critic_ablation} presents the learning curves (left-side) during training for centralized, per-class and per-agent critic architectures (shown in Figure~\ref{fig:critic}) in the heterogeneous PCP domain. Curves are showing the average number of steps taken across several batches ($\pm$ standard error) for each critic structure and once again, the fewer number of steps indicates a better approach. Moreover, the test results for policies learnt by each of the critic architectures are presented in Figure~\ref{fig:critic_ablation}, right-side. This figure shows the bar-plots representing the average number of steps taken to win the game by deploying the learnt policies through centralized, per-class and per-agent critic architectures. We see HetNet with per-class critic and HetNet with a per-agent critic have similar performance, both having better performance than fully-centralized HetNet, decreasing the number of steps of episode completion by \textcolor{black}{0.20} (10.01 $\to$ 9.81). The performance benefit can be attributed due to the ability to utilize individual rewards to some degree. Further investigation on these two variants is worthwhile and we leave it as a future direction.


\section{Conclusion and Future Work}
\label{sec:conclusion}
\noindent \textcolor{black}{We are motivated by} the problem of learning communication protocols among collaborating agents of a team that are of different types (e.g., class) and can have different state, observation and action spaces (i.e., heterogeneous agents). Without properly modeling such heterogeneity, communication can be detrimental since the agents do not ``speak" the same language (i.e., have different action spaces). 
We formulate our new problem as a Multi-Agent Heterogeneous Partially Observable Markov Decision Process (MAH-POMDP) and propose heterogeneous graph attention networks\textcolor{black}{, called HetNet,} to learn efficient and diverse communication models for coordinating agents \textcolor{black}{towards} accomplishing tasks that are of collaborative nature. \textcolor{black}{Specifically, we propose a Multi-Agent Heterogeneous Actor-Critic (MAHAC) learning paradigm to obtain collaborative per-class policies and effective communication protocols for a composite team such as perception-action robot teams~\cite{seraj2020firecommander}. We evaluate HetNet against two state-of-the-art, end-to-end communication learning baselines, namely CommNet~\cite{sukhbaatar2016learning} and IC3Net~\cite{singh2018learning} in both an adopted homogeneous environment (e.g., the Predator-Prey~\cite{singh2018learning}) and a heterogeneous environment (e.g., the Predator-Capture-Prey).} Our results validate the utility of HetNet in learning heterogeneous communication protocols for composite robot teams by demonstrating general applicability and higher performance than the baselines in both homogeneous and heterogeneous environments.

\textcolor{black}{In future work,} we intend to \textcolor{black}{stabilize the critic structures for further performance gain and better robustness} and to extend HetNet's utility to other multi-agent, heterogeneous domains with composite teams and collaborative tasks. We also intend to compare HetNet against more baselines including DIAL~\cite{foerster2016learning}, TarMAC~\cite{das2018tarmac} and MAGIC~\cite{niu2021multi}. Finally, we plan to design and incorporate class-specific encoder-decoder networks to the communication channel in HetNet to enable learning a shared ``language" among heterogeneous agents of a team.


\section*{Acknowledgments}
Supported by the Laboratory Directed Research and Development program at Sandia National Laboratories, a multimission laboratory managed and operated by National Technology and Engineering Solutions of Sandia LLC, a wholly owned subsidiary of Honeywell International Inc. for the U.S. Department of Energy’s National Nuclear Security Administration under contract DE-NA0003525. This paper describes objective technical results and analysis. Any subjective views or opinions that might be expressed in the paper do not necessarily represent the views of the U.S. Department of Energy or the United States Government.



\bibliographystyle{plainnat}  
\bibliography{main}

\end{document}